\documentclass[aip,graphicx]{revtex4-1}

\usepackage{lineno}
\usepackage{amsmath,amsfonts}
\usepackage{amssymb}
\usepackage{graphicx}
\usepackage{subcaption}
\usepackage{caption}
\usepackage{svg}
\usepackage{dcolumn}
\usepackage{bm}
\usepackage{comment}

\usepackage[utf8]{inputenc}
\usepackage[T1]{fontenc}
\usepackage{mathptmx}
\usepackage{etoolbox}
\renewcommand{\selectlanguage}[1]{}
\begin{document}


\title{Micro-Transfer Printing of Lithium Niobate on 200 mm Silicon Photonics: A High-Speed Heterogeneous Wafer-Scale Platform} 



\author{Xiujun Zheng}
\email[]{xiujun.zheng@ugent.be}
\affiliation{%
Department of Information Technology (INTEC) - Photonics Research Group, Ghent University–imec, Technologiepark Zwijnaarde 126, 9052 Ghent, Belgium%
}
\affiliation{%
imec, Kapeldreef 75, 3001 Leuven, Belgium%
}

\author{Suzanne Bisschop}
\affiliation{%
Department of Information Technology (INTEC) - Photonics Research Group, Ghent University–imec, Technologiepark Zwijnaarde 126, 9052 Ghent, Belgium%
}
\affiliation{%
imec, Kapeldreef 75, 3001 Leuven, Belgium%
}

\author{Arno Moerman}
\affiliation{%
Department of Information Technology (INTEC) - Photonics Research Group, Ghent University–imec, Technologiepark Zwijnaarde 126, 9052 Ghent, Belgium%
}
\affiliation{%
imec, Kapeldreef 75, 3001 Leuven, Belgium%
}

\author{Margot Niels}
\affiliation{%
Department of Information Technology (INTEC) - Photonics Research Group, Ghent University–imec, Technologiepark Zwijnaarde 126, 9052 Ghent, Belgium%
}
\affiliation{%
imec, Kapeldreef 75, 3001 Leuven, Belgium%
}

\author{Ewoud Vissers}
\affiliation{%
Department of Information Technology (INTEC) - Photonics Research Group, Ghent University–imec, Technologiepark Zwijnaarde 126, 9052 Ghent, Belgium%
}
\affiliation{%
imec, Kapeldreef 75, 3001 Leuven, Belgium%
}

\author{Athina Papadopoulou}
\affiliation{%
Department of Information Technology (INTEC) - Photonics Research Group, Ghent University–imec, Technologiepark Zwijnaarde 126, 9052 Ghent, Belgium%
}
\affiliation{%
imec, Kapeldreef 75, 3001 Leuven, Belgium%
}

\author{Philip Ekkels}
\affiliation{%
Department of Information Technology (INTEC) - Photonics Research Group, Ghent University–imec, Technologiepark Zwijnaarde 126, 9052 Ghent, Belgium%
}
\affiliation{%
imec, Kapeldreef 75, 3001 Leuven, Belgium%
}

\author{Patrick Nenezic}
\affiliation{%
Department of Information Technology (INTEC) - Photonics Research Group, Ghent University–imec, Technologiepark Zwijnaarde 126, 9052 Ghent, Belgium%
}
\affiliation{%
imec, Kapeldreef 75, 3001 Leuven, Belgium%
}

\author{Simone Atzeni}
\affiliation{%
Department of Information Technology (INTEC) - Photonics Research Group, Ghent University–imec, Technologiepark Zwijnaarde 126, 9052 Ghent, Belgium%
}
\affiliation{%
imec, Kapeldreef 75, 3001 Leuven, Belgium%
}

\author{Elif Ozceri }
\affiliation{%
Department of Information Technology (INTEC) - Photonics Research Group, Ghent University–imec, Technologiepark Zwijnaarde 126, 9052 Ghent, Belgium%
}
\affiliation{%
imec, Kapeldreef 75, 3001 Leuven, Belgium%
}

\author{Tiernan McCaughery }
\affiliation{%
Department of Information Technology (INTEC) - Photonics Research Group, Ghent University–imec, Technologiepark Zwijnaarde 126, 9052 Ghent, Belgium%
}
\affiliation{%
imec, Kapeldreef 75, 3001 Leuven, Belgium%
}

\author{Ali Uzun}
\affiliation{%
Department of Information Technology (INTEC) - Photonics Research Group, Ghent University–imec, Technologiepark Zwijnaarde 126, 9052 Ghent, Belgium%
}
\affiliation{%
imec, Kapeldreef 75, 3001 Leuven, Belgium%
}

\author{Ye Chen}
\affiliation{%
Department of Information Technology (INTEC) - Photonics Research Group, Ghent University–imec, Technologiepark Zwijnaarde 126, 9052 Ghent, Belgium%
}
\affiliation{%
imec, Kapeldreef 75, 3001 Leuven, Belgium%
}

\author{Laurens Bogaert}
\affiliation{%
Department of Information Technology (INTEC) - Photonics Research Group, Ghent University–imec, Technologiepark Zwijnaarde 126, 9052 Ghent, Belgium%
}
\affiliation{%
imec, Kapeldreef 75, 3001 Leuven, Belgium%
}

\author{Nishant Singh}
\affiliation{%
Department of Information Technology (INTEC) -  IDLab, Ghent University–imec, Technologiepark Zwijnaarde 126, 9052 Ghent, Belgium%
}
\affiliation{%
imec, Kapeldreef 75, 3001 Leuven, Belgium%
}

\author{Sandeep Seema Saseendran}
\affiliation{%
imec, Kapeldreef 75, 3001 Leuven, Belgium%
}

\author{Sofie Janssen}

\affiliation{%
imec, Kapeldreef 75, 3001 Leuven, Belgium%
}

\author{Natarajan Rajasekaran}
\affiliation{%
imec, Kapeldreef 75, 3001 Leuven, Belgium%
}
\author{Sadhishkumar Balakrishnan}
\affiliation{%
imec, Kapeldreef 75, 3001 Leuven, Belgium%
}

\author{Philippe Absil}
\affiliation{%
imec, Kapeldreef 75, 3001 Leuven, Belgium%
}

\author{Günther Roelkens}
\affiliation{%
Department of Information Technology (INTEC) - Photonics Research Group, Ghent University–imec, Technologiepark Zwijnaarde 126, 9052 Ghent, Belgium%
}
\affiliation{%
imec, Kapeldreef 75, 3001 Leuven, Belgium%
}

\author{Bart Kuyken}
\affiliation{%
Department of Information Technology (INTEC) - Photonics Research Group, Ghent University–imec, Technologiepark Zwijnaarde 126, 9052 Ghent, Belgium%
}
\affiliation{%
imec, Kapeldreef 75, 3001 Leuven, Belgium%
}

\author{Sarah Uvin}
\affiliation{%
Department of Information Technology (INTEC) - Photonics Research Group, Ghent University–imec, Technologiepark Zwijnaarde 126, 9052 Ghent, Belgium%
}
\affiliation{%
imec, Kapeldreef 75, 3001 Leuven, Belgium%
}

\author{Maximilien Billet}
\affiliation{%
Department of Information Technology (INTEC) - Photonics Research Group, Ghent University–imec, Technologiepark Zwijnaarde 126, 9052 Ghent, Belgium%
}
\affiliation{%
imec, Kapeldreef 75, 3001 Leuven, Belgium%
}


\date{\today}

\begin{abstract}
The rapid growth of artificial intelligence (AI) and other data-center applications is driving the demand for photonic interconnects that combine high-speed with low energy consumption, making scalability a critical requirement. Micro-transfer printing (MTP) has emerged as a promising technique for the wafer-scale heterogeneous integration of thin-film lithium niobate (TFLN) onto silicon photonics (SiPho) platforms. Here, we demonstrate heterogeneous SiPho–TFLN integration across four full 200 mm wafers with a 3$\sigma$ placement accuracy down to 420 nm and a printing yield of >95\%. Low insertion loss <2 dB over 600 phase modulators (forming 300 amplitude modulators) is achieved. A half-wave voltage of 4 V in push-pull configuration, and high-speed modulation with a bandwidth >70 GHz are demonstrated on a subset of tested devices. 
\end{abstract}


\maketitle 

\section{Introduction}
\label{sec:intro}  
The growing demand for low energy consumption, dense integration and fast components in data centers is pushing the performance requirements for photonic integrated circuits, and especially for optical modulators that enable optical data-communication links\cite{torrijos-moran_industry_2026}. Emerging AI and data-center interconnect roadmaps are pushing per-lane data rates toward hundreds of Gbit/s, placing increasing pressure on modulator bandwidth, drive voltage, linearity and energy efficiency. Conventional depletion-based silicon (Si) modulators face trade-offs in these metrics, motivating the exploration of new solutions. As an alternative, several technologies are considered to reach this milestone, such as the use of BTO \cite{chelladurai_barium_2024}, plasmonics \cite{kohli_plasmonic_2025,baeuerle_120_2019,eppenberger_resonant_2023}, graphene, \cite{wu_graphene-based_2024,rahimi_graphene-based_2026}, electro-absorption modulators \cite{steckler_monolithic_2025} and electro-optics materials. TFLN has emerged as a promising platform for high-performance modulators due to its intrinsically low optical loss, large electro-optic coefficient, and capability to support high-speed operations \cite{zhang_integrated_2021}. It is therefore seen as a solution to surpass the performance limitations of conventional SiPho platforms \cite{zhou_silicon_2024}. However, TFLN is not compatible with CMOS fabrication, primarily because of lithium contamination concerns within silicon foundries, which limits circuit complexity and the production volume of TFLN-based fabrication processes\cite{wandesleben_influences_2024}. 

To overcome these limitations, heterogeneous integration offers a viable approach for incorporating TFLN onto SiPho, combining the advantages of both materials on the same platform. Wafer-to-wafer and die-to-wafer bonding have produced high-performance SiPho-TFLN platforms and remain important integration routes \cite{ghosh_wafer-scale_2023,churaev_heterogeneously_2023,rahman_integration_2026,wu_heterogeneous_2025}. This method, demonstrated in the literature, provides platforms that can reach to the standards of the optical data-communication\cite{declercq_320_2025}. 
 However, blanket bonding consumes TFLN over the full bonded area, offers limited selectivity, and generally imposes strong constraints on target-wafer planarity and process sequencing.  To address this weakness, micro-transfer printing (MTP) has been evaluated \cite{roelkens_present_2024,yu_advancements_2025,niels_advances_2026} and provides a complementary route in which processed TFLN coupons are placed only at the required locations on a pre-fabricated silicon photonics wafer. Furthermore, the integration in local cladding apertures offers the possibility of the use of complex advanced silicon stacks  In this work, we demonstrate the wafer-scale integration of pre-patterned TFLN modulators on full 200-mm silicon photonics wafers using MTP. A total of 4 wafers with more than 600 phase modulators, corresponding to 300 amplitude modulators, are successfully transferred and integrated across the SiPho wafers. The fabricated heterogeneous devices are 7-mm-long push-pull Mach–Zehnder modulators (MZMs) in ground-signal-signal-ground (GSSG) electrode configuration. This demonstration highlights the scalability and the potential for volume production of high-speed SiPho-TFLN heterogeneous modulators fabricated with MTP.

\begin{figure*} [ht]
\begin{center}
\begin{tabular}{c} 
\includegraphics[width=\linewidth]{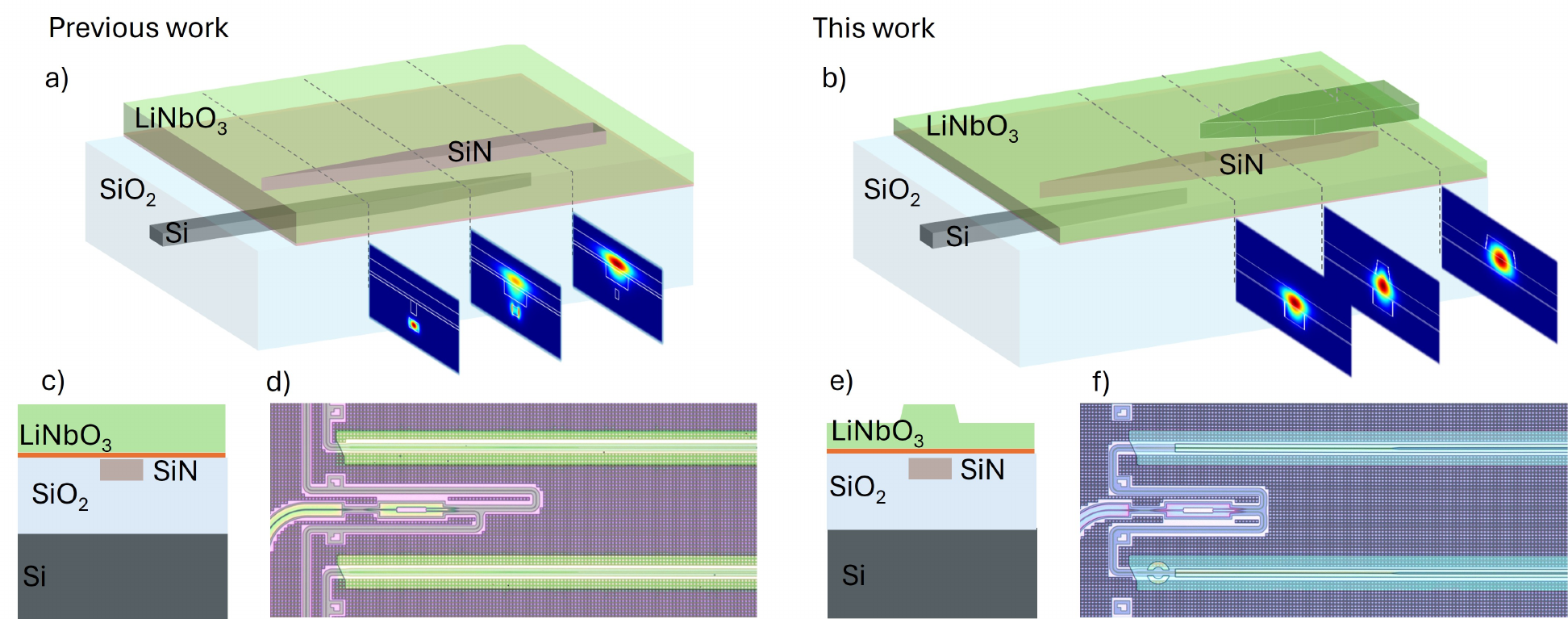}
\end{tabular}
\end{center}
\caption
{Comparison between previously reported designs and the optimised design presented in this work. a) The transfer printed TFLN device consists in a material slab, not allowing for optimised confinement of the light. b) The transfer printed devices in this work are patterned prior to the integration, enabling an optimised mode confinement. c) Cross-section of the design from a). d) Microscope picture of a chip made following the design from a). e) Cross-section of the design from b). f) Microscope picture of a chip made following the design from b).}
\label{fig:printing_overview}
\end{figure*} 

First, a statistical analysis of wafer-scale (200-mm) integration of patterned TFLN devices (including a TFLN waveguide), demonstrating high transfer yield and low optical insertion loss, are presented. As a proof of concept, back-end processing of metal electrodes is performed on a subset of the transferred devices, enabling reproducible fabrication of electro-optic modulators with low V$_{\pi}$. The radio-frequency (RF) performance of representative devices are also characterised.

\section{Heterogeneous thin-film lithium niobate on silicon photonics using micro-transfer printing}

MTP is an emerging method for heterogeneous integration. The technology, licensed from X-Celeprint Ltd., combines die-level assembly with wafer-scale processing. Thin-film devices (“coupons”) are first fabricated in dense arrays on a source wafer. They are released by selectively etching away a sacrificial layer. Once suspended, an elastomeric stamp, generally made out
of polydimethylsiloxane (PDMS), is used to retrieve the devices. The coupons are printed onto a target wafer and bonded using adhesive or direct bonding.
MTP offers high material efficiency, known-good-die integration, and excellent scalability through parallel printing, enabling high throughput at sub-micron alignment accuracy. Its greatest strength lies in its broad material and process compatibility, allowing heterogeneous integration of multiple material systems without compromising individual fabrication flows. The integration is performed at the back-end of SiPho processing, supporting both efficient optical coupling and electrical interconnection via redistribution layers.

Despite these advantages, MTP remains at an early stage of commercialisation, with challenges related to pritning yield, long-term reliability, and supply-chain maturity. Nonetheless, extensive research demonstrations including the integration of III-V lasers, amplifiers, modulators, photo-diodes, and thin-film electro-optic devices highlight MTP’s strong potential for enabling complex, high-performance heterogeneous photonic integrated circuits\cite{chen2026micro}.

\section{Demonstration of integration of TFLN on 200-mm silicon photonics wafers}\label{sec:Demonstration of integration of TFLN on 200-mm SiPho wafers}

The modulator demonstrated in this work is a hybrid SiN/TFLN unbalanced Mach–Zehnder modulator (MZM), where silicon waveguides are used for routing and SiN/TFLN hybrid waveguides form the electro-optic modulating arms. The MZM consists of TFLN electro-optically active arms combined with a passive silicon photonic circuit. The passive components, including grating couplers, Si and SiN waveguides and multi-mode interferometer (MMI) splitters, are designed using a standard process design kit (PDK) and fully fabricated prior to active integration. The TFLN devices are prepared and suspended before being integrated onto the silicon photonics wafer. This integration is done using micro-transfer printing with a 50-nm intermediate bonding layer. During transfer printing, the LN crystal orientation in the two modulator arms is intentionally flipped, enabling push–pull modulation. Optical coupling from Si waveguides into the hybrid SiN/TFLN modes is achieved through a bilayer Si–SiN adiabatic transition, rendering a low-loss transition. Electro-optic modulation is provided by the Pockels effect in TFLN, while the SiN layer ensures low-loss optical propagation. Finally, metal electrodes are defined in a post-processing step.
Fig.\ref{fig:printing_overview} compares previously reported device designs \cite{niels2025demonstration} and the upgraded design. In previous designs, as illustrated in Fig.\ref{fig:printing_overview} a) the TFLN is a slab and the mode is a hybrid SiN/TFLN mode. In contrast, the current design incorporates a TFLN taper and an etched waveguide, as seen in Fig.\ref{fig:printing_overview} b). The optical mode confinement is stronger, allowing for closer electrode spacing. This enables for a lower V$_{\pi}$, better optical transition and improved mode confinement along the propagation direction. Furthermore, this concept is compatible with designs asking for full coupling in a TFLN waveguide and is not limited to SiN/TFLN hybrid modes, even if this was not demonstrated in this work. Details on the cross-sections of both designs and the corresponding fabricated modulators are shown in Fig.\ref{fig:printing_overview}(c-f).

\begin{figure*} [ht]
\begin{center}
\begin{tabular}{c} 
\includegraphics[width=\linewidth]{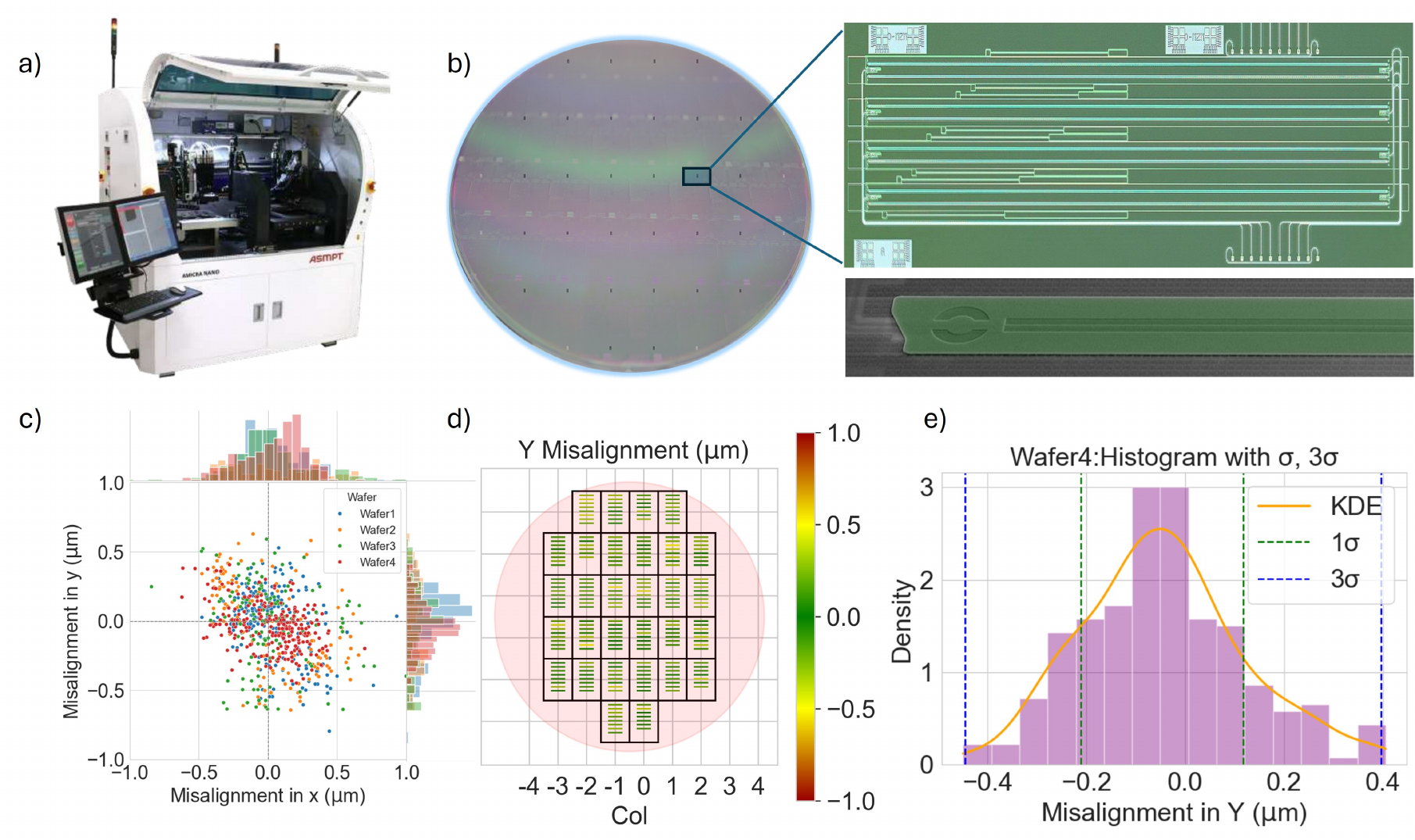}
\end{tabular}
\end{center}
\caption
{Overview of the integration of TFLN on a 200-mm SiPho wafer using MTP. a) Wafer-scale printing tool compatible with target wafers of 200 mm and 300 mm. b) Overview of a 200-mm SiPho wafer, which is completely populated with TFLN and zoomed images of a reticle and a TFLN printed device. c) Statistical printing alignment data, extracted from four wafers. d) Distribution wafer map of Y misalignment data across a wafer. The distribution of the alignment focuses on the vertical (y) direction since this is the relevant metric for low loss optical coupling e) Gaussian Kernel Density Estimation (KDE) distribution data for one wafer}
\label{fig:wave-scale with micra transfer printing}
\end{figure*} 

\newpage
The approach presented in this work is compatible with the integration of TFLN on 200-mm (and 300-mm) wafers using a commercial MTP tool, the ASMPT Amicra NANO, as depicted in Fig.\ref{fig:wave-scale with micra transfer printing} a). The used approach enables accurate and repeatable wafer-scale transfer of TFLN  from a source wafer (providing the TFLN suspended coupons) to a target wafer (the SiPho platform). The entire process is enabled by a fully automated tool that integrates coupon tracking, stamp identification, printing location control, and misalignment pattern recognition. Each full cycle duration is around 1 min, for single coupon printing operation as well as for the transfer of arrays of multiple devices, including the coupon recognition, location to print identification, picking, alignment, printing, post-printing measurement and stamp cleaning. Demonstrations up to 28 coupons of 1-mm \cite{niels2025demonstration} and 16 coupons of 7-mm have been performed, where the printing duration and alignment level were confirmed. Here we provide data by single coupon printing on four 200-mm SiPho wafers. An example of a chip containing four MZMs after TFLN integration on the wafer is presented in Fig.\ref{fig:wave-scale with micra transfer printing} b). The SEM picture highlights the patterned TFLN waveguide. All back-end of line processing following the CMOS wafer production has been implemented in the imec-Gent University pilot line "TRANSVERSE"\cite{Transverse}.

A dedicated full-wafer calibration is required to capture systematic tool offsets and stabilize the tool, thereby allowing global optimization of the tool settings. Wafer No.~1 was used as a calibration wafer to set picking and printing recipes and evaluate alignment accuracy across all coupons. The alignment 3$\sigma$ value was obtained from automated post-printing measurements provided by the tool. Additional microscope inspection was carried out on randomly chosen devices to validate the results. The readout alignment of the commercial transfer-printing tool was not optimised and Wafer No.~1 has been used as a printing test platform (3$\sigma \approx 770$ nm) to confirm the stability of the printing operation. This wafer also demonstrates the minimum effort to set a new printing procedure. In total, four successive wafers are printed using a similar printing recipe architecture. As shown in Fig.\ref{fig:wave-scale with micra transfer printing} c), the misalignment distributions along both the x and y axes follow Gaussian profiles centred at zero. Table~\ref{tab:misalignment table} summarises the extracted $\sigma$ and $3\sigma$ values for the four processed wafers. Progressive improvement in the $\sigma$ values is observed as the printing operation is refined by playing with the pattern recognition parameters, to finally reach the representative value given by wafer No.~4 exhibiting a value of 3$\sigma$ below 500 nm. Fig.\ref{fig:wave-scale with micra transfer printing} d) shows the spatial distribution of alignment error for the y-direction (the most critical for the coupling) across a representative wafer, demonstrating uniform printing performance over the full wafer area and Fig.\ref{fig:wave-scale with micra transfer printing} e) shows the Gaussian distribution. In total, 600 pre-patterned coupons, incorporating TFLN waveguides, were printed with a high transfer yield of >95\% across all wafers, with the yield defined as the number of successfully printed coupons without any breakage. The failures can be explained by remaining inhomogeneities in the source and target preparation. The printing process could reach a higher yield by the implementation of automatic inspection in all preparation steps, hence utilising the know-good-die concept attributed to MTP. These results demonstrate stable wafer-scale coupon preparation and printing. Implementation in industrial fabrication environments would enable larger statistical datasets and further validation of the long-term process stability.

\begin{table*}
\caption{3$\sigma$ data in y axis for 4 wafers populated with TFLN with MTP}
\label{tab:misalignment table}
\centering
\begin{ruledtabular}
\begin{tabular}{ccccc}
Wafer & Printing success$/$printing trials  & Printing yield &$\sigma$ in Y (nm)  & 3$\sigma$ in Y (nm)  \\
\hline
Wafer 1 &  260/272 & 96\% & 259 & 770 \\
\hline
Wafer 2 &  108/114 &95\%& 324 & 630 \\
\hline
Wafer 3 &  145/152 &95\%& 289 & 610 \\
\hline
Wafer 4 &  234/240 &98\%& 165 & 420 \\
\end{tabular}
\end{ruledtabular}
\end{table*}

\section{Optical characterisation of the devices}

The optical insertion loss (IL) of each device across the wafers must be characterised to evaluate process uniformity and device yield. The ILs are excluding the silicon routing from the CMOS platform as described in section \ref{sec:Demonstration of integration of TFLN on 200-mm SiPho wafers}. As photonic integration scales to larger wafer formats and higher device densities, device variability becomes an increasingly important consideration. High-throughput measurement capabilities are therefore required to generate large datasets that provide statistical insight into device performance and process uniformity.

\begin{figure*} [ht]
\begin{center}
\begin{tabular}{c} 
\includegraphics[width=\linewidth]{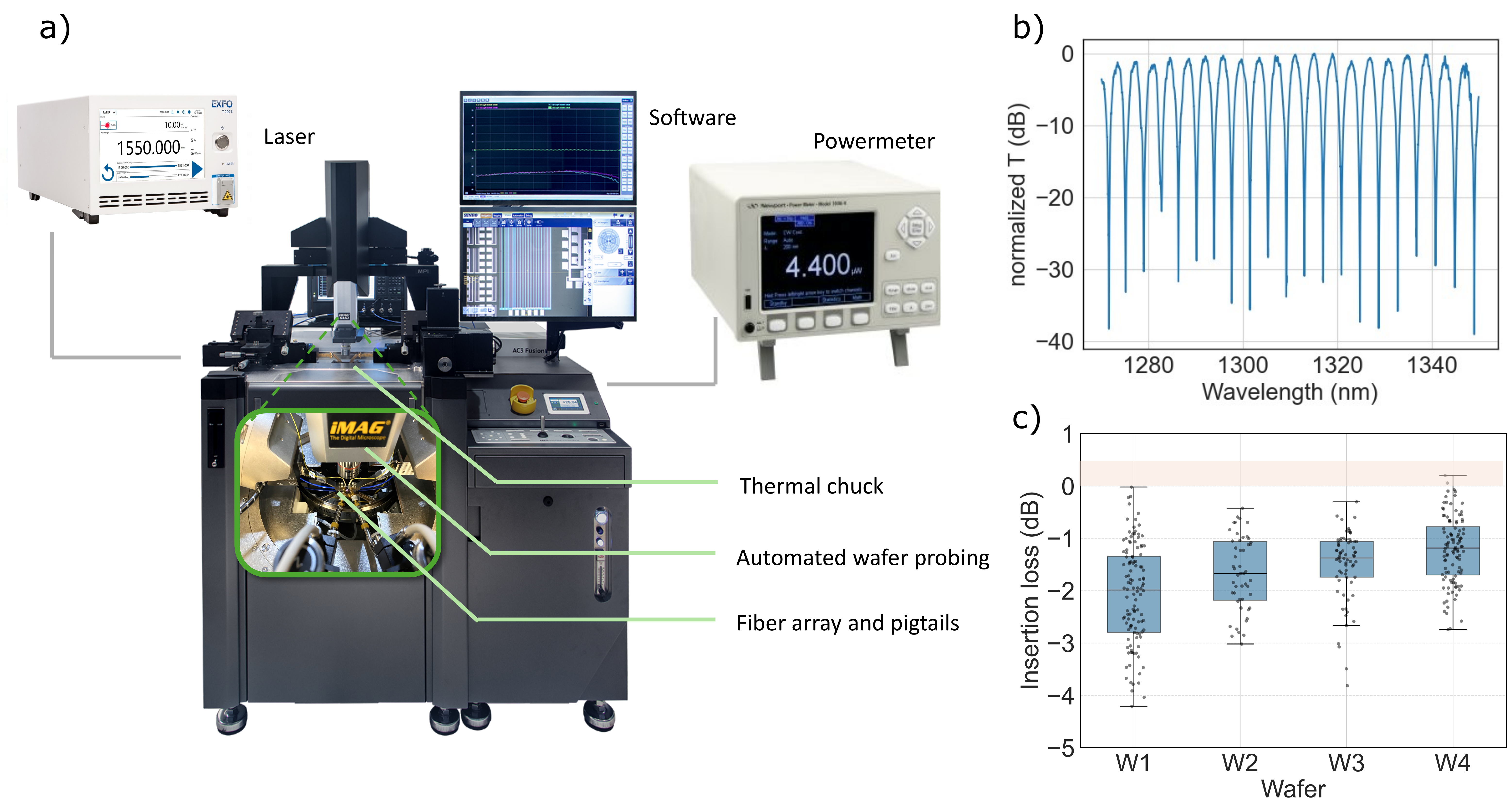}
\end{tabular}
\end{center}
\caption
{ a) Characterisation setup for 200-mm wafers, based on an automated wafer prober.  b) Example of a normalised MZM transmission spectrum.  c) Statistical insertion loss of all four wafers.}
\label{fig:MPI and example of IL data}
\end{figure*}

A wafer-scale characterisation setup, based on an MPI TS2000-IFE fully automated probe station, was used, where the optical characterisation was performed using an EXFO CTP10 testing platform in combination with an EXFO T200S tunable laser source and an optical power meter module (1936-R Newport). A schematic representation of the setup is provided in Fig.~\ref{fig:MPI and example of IL data} a). The measurement setup enables fast and repeatable characterisation of photonic devices across an entire wafer, mandatory for wafer level characterisation. An example of a wavelength sweep around 1310 nm performed on a representative MZM prior to metallisation, is shown in Fig.~\ref{fig:MPI and example of IL data} b). The measured transmission spectrum is normalised and excludes the grating coupler and device ILs which are not contributing to the integrated phase shifters. The extinction ratio exceeding 30 dB for this device demonstrated the quality of the MZM interferometer.

The ILs are measured for all printed modulators on the four wafers and are presented in Fig.~\ref{fig:MPI and example of IL data} c). To obtain the ILs, a reference waveguide with a given length is first measured, after which the corresponding MZMs are characterised. The observed spread in the mean value of the ILs for each wafer is correlated with the misalignment in table \ref{tab:misalignment table}. An increase in the printing alignment accuracy (lower 3$\sigma$) provides lower ILs. The spread of the IL values within a given wafer is attributed to multiple factors, primarily coupon misalignment and measurement uncertainty. This uncertainty is partly explained by the difficulty in extracting very low ILs from short devices (7-mm long). All the devices presented here provide decent ILs, with a mean value below 2 dB. By using separated test structures based on ring resonators, the precision of the measurements could be improved.

\section{Electro-optic characterisation of the devices}

\begin{figure*} [ht]
\begin{center}
\begin{tabular}{c} 
\includegraphics[width=\linewidth]{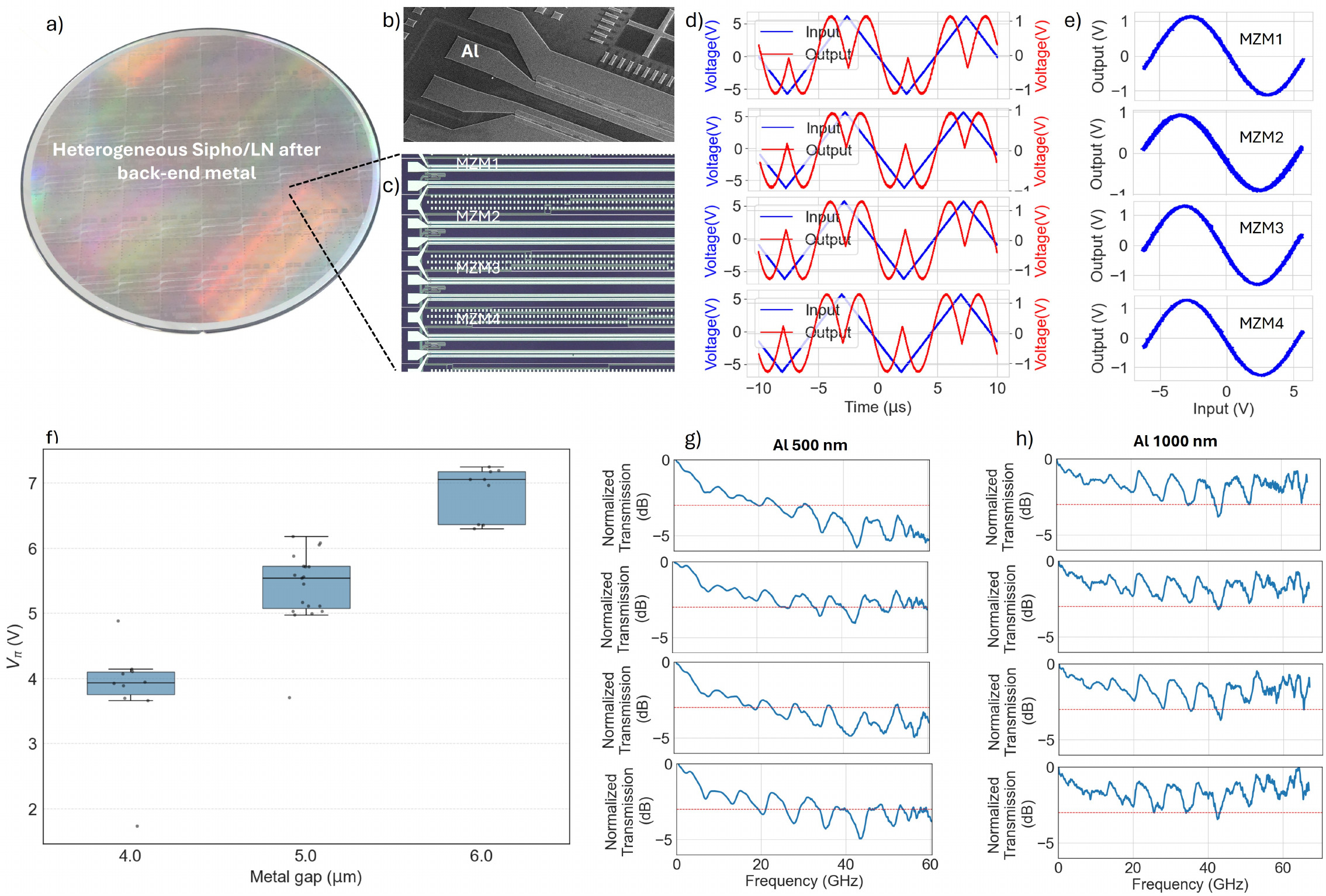}
\end{tabular}
\end{center}
\caption
{Electro-optics characterisation. a) Picture of a 200-mm wafer after RDL back-end processing.  b-c) Zoomed pictures on a device and on a chip. d) RF and photodiode signal as a function of time.  e) V$_{\pi}$ data analysis.  f) V$_{\pi}$ as a function of the electrode gaps.  g-h) EOE measurement of four MZMs on a chip for 500~nm and 1~\textmu m thick Al electrodes}
\label{fig:EOE measurement of 4 MZM}
\end{figure*}

In addition to passive optical characterisation, wafer-level V$_{\pi}$ measurements were also performed to evaluate the electro-optic modulation efficiency of the transferred TFLN devices.
The back-end redistribution layer (RDL) processing was carried out on a full 200-mm wafer using UV-lithography, metal deposition and lift-off. This study is limited to one wafer because of the facility constraints. The value of the electrode gap design varies between 4 and 6~\textmu m and aluminium (Al) electrodes with a thickness of 500~nm are used. A thin layer of 100 nm oxide is placed in between the RDL and the TFLN as a spacer. Fig.~\ref{fig:EOE measurement of 4 MZM} a) depicts the wafer after the fabrication of the electrodes, while Fig.~\ref{fig:EOE measurement of 4 MZM} b-c) presents an SEM image and an optical microscope picture of the devices.  \\
The devices are driven by an RF signal generator, providing a triangular wave at a frequency of 100~kHz. The optical signal is modulated via the EO effect and recorded by a photodetector. The raw signals of four MZMs from the same chip are presented in Fig.~\ref{fig:EOE measurement of 4 MZM} d), where the signal of both the RF signal generator and the photodetector are presented, as a function of time. The four devices show very similar behavior. In order to extract the V$_{\pi}$ of the amplitude modulators, the photodiode signal is plotted as a function of the RF signal for a full modulation period. The analysis of the data from Fig.~\ref{fig:EOE measurement of 4 MZM} d) is presented in Fig.~\ref{fig:EOE measurement of 4 MZM} e). In this case, the V$_{\pi}$ is around 4~V and corresponds to an electrode gap of 4~\textmu m. Results of the V$_{\pi}$ value as a function of the electrode gap are also presented in Fig.~\ref{fig:EOE measurement of 4 MZM} f). As expected from their linear relationship, a wider electrode gap results in a higher V$_{\pi}$, confirming the linear dependence between these two parameters. However, with the current technology used to process the RDLs, the electrode misalignment is around 1~\textmu m, making it difficult to stay at low ILs after metalisation. On average, an extra 2~dB IL is observed at this step. The implementation of alternative RDL processing approaches (use of a stepper, printing TFLN devices with pre-defined electrodes or using buried electrodes \cite{boynton_heterogeneously_2020}) could help reduce the additional IL while further decreasing the electrode gap and the V$_{\pi}$. Nevertheless, this proof of concept shows the possibility of making active EO devices based on MTP at a wafer level.

In this platform, the substrate is made with a high resistivity silicon, allowing for very high speed operations (>100 GHz) and this property is evaluated on a subset of devices.  For this experiment, the RDLs are fabricated on two extra chips. The deposited Al thickness is 500~nm on the first one and 1~\textmu m on the second. While the V$_{\pi}$ is negligibly affected by the value of the metal thickness, the electric to optic to electric (EOE) bandwidth (BW) differs. As observed in Fig.~\ref{fig:EOE measurement of 4 MZM} g-h), the bandwidth is limited to 20-30 GHz using a thin 500 nm thick RDL, while the BW is extended to 70 GHz+ for the option with 1~\textmu m Al. The exact BW is expected to be around 90 GHz, but the current experiment is limited by the measurement tools. Further design of the device architecture would allow for improvement in the performance, as presented in the work of P. Nenezic et al.\cite{nenezic2026variability}, taking into account realistic process variability stemming from experimental statistical analysis, extracted from the presented results.

\section{Conclusion}
This work demonstrates the scalability of micro-transfer printing for wafer-scale integration of TFLN modulators on 200-mm silicon photonics platforms. A printing yield of >95\% and a 3-sigma alignment accuracy below 500 nm enable efficient optical coupling, as evidenced by an insertion loss lower than 2~dB, demonstrated over 4 wafers. Wafer-scale measurements show good uniformity across the wafers, confirming the robustness of the integration process. Wafer-level half-wave voltage characterisation further validates the electro-optic performance and compatibility with automated testing. The EO effect provided by TFLN is not affected by the transfer printing process, and devices with 70 GHz+ BW, made from a subset of fabricated wafers, are demonstrated. Overall, the results highlight micro-transfer printing as a viable approach for high-volume integration of high-performance TFLN high-speed modulators within CMOS-compatible silicon photonics platforms. Recent demonstrations of the integration of CMOS electronics circuits (EIC) using MTP \cite{li_3d-integrated_2026} and co-integration of heterogeneous MZMs with electronic drivers and transimpedance amplifiers\cite{declercq_320_2025}, are paving the way for the next generation of optical interconnects using the presented technology. Furthermore, several demonstrations of heterogeneous integration based on the use of lithium tantalate as an alternative material have been demonstrated \cite{cai_heterogeneously_2026} . The proposed methodology based on MTP is also compatible with this material \cite{niels_high-speed_2026, su_low-loss_2025}, allowing for high-power applications or working with short wavelengths down to the UV-range\cite{lin_thin-film_2026}.

\begin{acknowledgments}
We would like to thank ASMPT Amicra team for their valuable input for the tool utilisation. The authors would like to also thank the imec-Leuven teams providing the silicon and silicon nitride photonic waveguide circuits for our pilot line. The research has been made possible by FWO and F.R.S.-FNRS under the Excellence of Science
(EOS) program (40007560). The work has been supported by The Dutch National Growth Fund PhotonDelta, INTERREG Vlaanderen-Nederland project LIGHTUP, and CHIPS-JU PhotonixFAB (101111896). 
\end{acknowledgments}


%
%

%


\bibliography{sample}

\end{document}